%Paper: hep-ph/9408367
%From: cliff@hep.Physics.McGill.CA (Cliff Burgess)
%Date: Thu, 25 Aug 94 18:27:53 +0200
%Date (revised): Wed, 12 Oct 94 21:05:24 +0200

% Required Macros -------------------------------------------------

% ----------------------------------------------------------------------
% Size and shape
% -----------------------------------------------------------------------

\font\titlefont = cmr10 scaled\magstep 4
\font\sectionfont = cmr10
\font\littlefont = cmr5% for equation names in draftmode
\font\eightrm = cmr8 

\def\ss{\scriptstyle} 
\def\sss{\scriptscriptstyle} 

\newcount\tcflag
\tcflag = 0  % this flag indicates whether harvmac is to be used
 %turn off 12 point definition if run with harvmac

\ifnum\tcflag = 0 \magnification = 1200 \fi  % 12 point

%\global\hsize = 5in% size of text
%\global\lmargin = 0.125in% 
\global\baselineskip = 1.2\baselineskip% line skip
\global\parskip = 4pt plus 0.3pt% paragraph skip
\global\abovedisplayskip = 18pt plus3pt minus9pt
\global\belowdisplayskip = 18pt plus3pt minus9pt
\global\abovedisplayshortskip = 6pt plus3pt
\global\belowdisplayshortskip = 6pt plus3pt

\def\barsoff{\overfullrule=0pt}

% -----------------------------------------------------------------------
% Draft mode stuff
% -----------------------------------------------------------------------

\def\endignore{}
\def\ignore #1\endignore{}% use to "comment out" text

\newcount\dflag% draft mode flag
\dflag = 0% initialize

% Time commands ---------------------------------------------------------

\def\monthname{\ifcase\month% for month numbers
\or January \or February \or March \or April \or May \or June%
\or July \or August \or September \or October \or November %
\or December% monthname 
\fi}

\newcount\dummy
\newcount\minute  % defines counters for the timestamp
\newcount\hour
\newcount\localtime
\newcount\localday
\localtime = \time
\localday = \day

\def\advanceclock#1#2{ % advances clock to give local time
\dummy = #1
\multiply\dummy by 60
\advance\dummy by #2
\advance\localtime by \dummy
\ifnum\localtime > 1440 % advances day if clock is advanced past midnight
\advance\localtime by -1440
\advance\localday by 1
\fi}

\def\settime{{\dummy = \localtime%
\divide\dummy by 60%
\hour = \dummy% hour
\minute = \localtime%
\multiply\dummy by 60%
\advance\minute by -\dummy% minutes
\ifnum\minute < 10 
\xdef\spacer{0} % leading 0 for minutes
\else \xdef\spacer{} 
\fi %
\ifnum\hour < 12 
\xdef\ampm{a.m.} % before noon
\else 
\xdef\ampm{p.m.} % after noon
\advance\hour by -12 %
\fi %
\ifnum\hour = 0 \hour = 12 \fi% make midnight, noon = 12
\xdef\timestring{\number\hour : \spacer \number\minute%
\thinspace \ampm}}}

% Draft mode commands ---------------------------------------------------

% -----------------------------------------------------------------------
% Headers
% -----------------------------------------------------------------------

\def\endtitle{}
\def\title#1\endtitle{\vskip.5in\titlefont
\global\baselineskip = 2\baselineskip% set line skip
#1\vskip.4in% title
\baselineskip = 0.5\baselineskip\rm}
 
\def\endauthors{}
\def\authors#1\endauthors{#1}

\def\endabstract{}
\def\abstract#1\endabstract{\vskip .3in%
\centerline{\sectionfont\bf Abstract}%
\vskip .1in
\noindent#1}

\def\nopageonenumber{\footline={\ifnum\pageno<2\hfil\else
\hss\tenrm\folio\hss\fi}}  % turns off page number on title page

\newcount\nsection% section counter
\newcount\nsubsection% subsection counter

% start new section
\def\section#1{\global\advance\nsection by 1% increment section number
\nsubsection=0
% section title
\bigskip\noindent\centerline{\sectionfont \bf \number\nsection.\ #1}
\bigskip\rm\nobreak}% back to normal

% start new subsection
\def\subsection#1{\global\advance\nsubsection by 1% increment subsection number
% subsection title
\bigskip\noindent\sectionfont \sl \number\nsection.\number\nsubsection)\
#1\bigskip\rm\nobreak}% back to normal

% unnumbered itemized topics 

% start new appendix
\def\appendix#1#2{\bigskip\noindent%
\centerline{\sectionfont \bf Appendix #1.\ #2}% appendix title
\bigskip\rm\nobreak}% back to normal

% -----------------------------------------------------------------------
% References
% -----------------------------------------------------------------------

\newcount\nref% create counter for references
\global\nref = 1% initialize it

% this is just in case there are no references at all but \listrefs
% is nevertheless used
\def\therefs{} 

 % puts the next reference on the next line

\def\ref#1#2{\xdef #1{[\number\nref]}% define reference label
\ifnum\nref = 1\global\xdef\therefs{\item{[\number\nref]} #2\ }% first ref
\else% not the first ref
\global\xdef\oldrefs{\therefs}% old reference list
\global\xdef\therefs{\oldrefs\vskip.1in\item{[\number\nref]} #2\ }%
\fi%
\global\advance\nref by 1% advance label
}

\def\listrefs{\vfill\eject\section{References}\therefs}

% -----------------------------------------------------------------------
% Footnotes
% -----------------------------------------------------------------------

\newcount\nfoot% create counter for footnotes
\global\nfoot = 1% initialize it

\def\foot#1#2{\xdef #1{(\number\nfoot)}% define footnote label
\footnote{${}^{\number\nfoot}$}{\eightrm #2}
\global\advance\nfoot by 1% advance label
}

% -----------------------------------------------------------------------
% Figures
% -----------------------------------------------------------------------

\newcount\nfig% create counter for figures
\global\nfig = 1% initialize it
\def\thefigs{} % in case there are no figures but captions appears 

\def\figure#1#2{\xdef #1{(\number\nfig)}% define figure label
\ifnum\nfig = 1\global\xdef\thefigs{\item{(\number\nfig)} #2\ }% first figure
\else% not the first figure
\global\xdef\oldfigs{\thefigs}% old figure caption list
\global\xdef\thefigs{\oldfigs\vskip.1in\item{(\number\nfig)} #2\ }%
\fi%
\global\advance\nfig by 1 } % advance label

\def\figurecaptions{\vfill\eject\section{Figure Captions}\thefigs}

% this is the old figure definition which is kept to
% keep the macros compatible with older papers
\def\fig#1{\xdef #1{(\number\nfig)}% define figure label
\global\advance\nfig by 1 } % advance label

% -----------------------------------------------------------------------
% Equations
% -----------------------------------------------------------------------

\newcount\cflag% create custom flag
\newcount\nequation% create equation counter
\global\nequation = 1% initialize it
\def\eqlabel{(1)}% initialize equation label

% Increment equation counter
\def\nexteqno{\ifnum\cflag = 0% if no custom numbering
\global\advance\nequation by 1% advance number
\fi% end of conditional
\global\cflag = 0% reset custom flag
\xdef\eqlabel{(\number\nequation)}}% define equation label

% Decrement equation counter
\def\lasteqno{\global\advance\nequation by -1% decrease number
\xdef\eqlabel{(\number\nequation)}}% define equation label

% Label equation
\def\label#1{\xdef #1{(\number\nequation)}% define equation name
\ifnum\dflag = 1% if in draft mode
{\escapechar = -1% locally remove "\"
\xdef\draftname{\littlefont\string#1}}% define draft name (small font)
%\xdef\draftname{\tt\string#1}}% define draft name (typewriter font)
\fi}

% Custom label equation
\def\clabel#1#2{\xdef\eqlabel{(\number\nequation #2)}% define custom label
\global\cflag = 1% set custom flag
\xdef #1{\eqlabel}% label equation
\ifnum\dflag = 1% if in draft mode
{\escapechar = -1% locally remove "\"
\xdef\draftname{\string#1}}% define draft name
\fi}

% Completely custom label equation
\def\cclabel#1#2{\xdef\eqlabel{#2)}% define custom label
\global\cflag = 1% set custom flag
\xdef #1{\eqlabel}% label equation
\ifnum\dflag = 1% if in draft mode
{\escapechar = -1% locally remove "\"
\xdef\draftname{\string#1}}% define draft name
\fi}

% Display equation stuff ------------------------------------------------

% End of equation
\def\eeq{}

% Begin displayed unnumbered equation
\def\eqnn #1\eeq{$$ #1 $$}

% Begin displayed numbered equation
\def\eq #1\eeq{
\ifnum\dflag = 0% if not in draft mode
{\xdef\draftname{\ }}% default = no draft name
\fi % end conditional
$$ #1% print equation
\eqno{\eqlabel \rlap{\ \draftname}} $$% print equation number
\nexteqno}% increment equation number

% Print equation number

% Equation array stuff --------------------------------------------------

% End line with equation number
% clear draft name

% Last eol
% clear draft name

% End line without equation number
% clear draft name

% Last eol without equation number
% clear draft name

% begin equation array
\def\eqa #1\eeq{
\ifnum\dflag = 0% if not in draft mode
{\xdef\draftname{\ }}% default = no draft name
\fi % end conditional
$$ \eqalignno{ #1 } $$% print equation
\global\cflag = 0}% reset custom flag

% -----------------------------------------------------------------------
% Useful abbreviations
% -----------------------------------------------------------------------

\def\ie{{\it i.e.\/}}

\def\etal{{\it et.al.\/}}

\def\via{{\it via\/}}

% -----------------------------------------------------------------------
% Journal abbreviations
% -----------------------------------------------------------------------

\def\ijmp#1#2#3{{\it Int.\ J.\ Mod.\ Phys.} {\bf A#1} (19#2) #3}

\def\npb#1#2#3{{\it Nucl.\ Phys.} {\bf B#1} (19#2) #3}
\def\plb#1#2#3{{\it Phys.\ Lett.} {\bf #1B} (19#2) #3}

\def\prd#1#2#3{{\it Phys.\ Rev.} {\bf D#1} (19#2) #3}

\def\prl#1#2#3{{\it Phys.\ Rev.\ Lett.} {\bf #1} (19#2) #3}

% Math parameters -----------------------------------------------------------

\global\nulldelimiterspace = 0pt

% Math relations ------------------------------------------------------------

% Math operations -----------------------------------------------------------

\def\frac#1#2{{{#1} \over {#2}}\,}  % fraction
\def\hf{{1\over 2}}

  % small fraction

  % derivative
  % partial derivative

\def\Dsl{\hbox{/\kern-.6700em\it D}} % D slash
\def\dsl{\hbox{/\kern-.5300em$\partial$}}
\def\pxpsl{\hbox{/\kern-.5600em$p$}}
\def\ssl{\hbox{/\kern-.5300em$s$}}
\def\epssl{\hbox{/\kern-.5100em$\epsilon$}}
\def\delsl{\hbox{/\kern-.6300em$\nabla$}}
\def\lxpsl{\hbox{/\kern-.4300em$l$}}
\def\elxpsl{\hbox{/\kern-.4500em$\ell$}}
\def\kxpsl{\hbox{/\kern-.5100em$k$}}
\def\qxpsl{\hbox{/\kern-.5000em$q$}}
 % transpose
\def\sla#1{\raise.15ex\hbox{$/$}\kern-.57em #1}% Feynman slash
\def\Pl{\gamma_{\sss L}}

% Math accents --------------------------------------------------------------

%\def\supsub#1#2{\mathstrut^{\vphantom{\dagger}#1}_{\vphantom{A}#2}}
%\def\sub#1{\mathstrut^{\vphantom{\dagger}}_{\vphantom{A}#1}}
%\def\sup#1{\mathstrut_{\vphantom{A}}^{\vphantom{\dagger}#1}}
%\def\rsub#1{\mathstrut^{\vphantom{\dagger}}_{\vphantom{A}\rm #1}}
%\def\rsup#1{\mathstrut_{\vphantom{A}}^{\vphantom{\dagger}\rm #1}}

\def\roughly#1{\mathrel{\raise.3ex\hbox{$#1$\kern-.75em\lower1ex\hbox{$\sim$}}}}
\def\lsim{\roughly<}
\def\gsim{\roughly>}

\def\tw#1{\tilde{#1}}
\def\ol#1{\overline{#1}}

% Alphabets ------------------------------------------------------------------

% Lower Case Bold Face

\def\bfp{{\bf p}}

\def\bfr{{\bf r}}

% Upper Case Bold Face

% Upper Case Script

\def\Scl{{\cal L}}

\def\Scw{{\cal W}}

% Upper Case ScriptScriptStyle

\def\ssa{{\sss A}}

\def\ssd{{\sss D}}

\def\ssf{{\sss F}}

\def\ssl{{\sss L}}

\def\ssn{{\sss N}}

\def\sst{{\sss T}}

\def\ssv{{\sss V}}

\def\ssz{{\sss Z}}

% Math functions -------------------------------------------------------------

% Math constructs ------------------------------------------------------------

% bras 'n' kets
\def\bra#1{\langle #1 |}
\def\ket#1{| #1 \rangle}

\def\Avg#1{\left\langle #1 \right\rangle}

% integral measures

% Abbreviations --------------------------------------------------------------

\def\hc{{\rm h.c.}}

% units

\def\eV{{\rm \ eV}}

\def\MeV{{\rm \ MeV}}
\def\GeV{{\rm \ GeV}}

% Local Definitions ---------------------------------------------------------

\def\Nbar{\ol{N}}
\def\Lbar{\ol{L}}

\def\npm{{\ssn_\pm}}
\def\np{{\ssn_+}}
\def\nm{{\ssn_-}}
\def\npmpm{{\ssn'_\pm}}
\def\npmp{{\ssn'_+}}
\def\npmm{{\ssn'_-}}
\def\lambdae{{\lambda_e}}
\def\twlambdae{{\tw{\lambda}_e}}

\def\bk{\item{}}
\def\bb{\beta \beta}
\def\bbzn{\beta \beta_{0\nu}}
\def\nue{\nu_e}

\def\cs#1{c_#1}
\def\sn#1{s_#1}
\def\csone{\cs{1}}
\def\cstwo{\cs{2}}

\def\snone{\sn{1}}
\def\sntwo{\sn{2}}

\def\GF{G_\ssf}
\def\pf{p_\ssf}

% Title page ----------------------------------------------------------------

\nopageonenumber
\baselineskip = 16pt
\barsoff

\rightline{August, 1994.}
\line{hep-ph/9408367 \hfil McGill-94/37, NEIP-94-007, UMD-PP-95-11}
%\vskip .1in
\title
\centerline{Heavy Sterile Neutrinos}
\centerline{and}
\centerline{Neutrinoless Double Beta Decay}
\endtitle
%\vskip 0.1in
\authors
\centerline{P. Bamert${}^a$, C.P. Burgess${}^{a,b}$ and R.N. Mohapatra${}^c$}
\vskip .15in
\centerline{\it ${}^a$ Institut de Physique, Universit\'e de Neuch\^atel}
\centerline{\it 1 Rue A.L. Breguet, CH-2000 Neuch\^atel, Switzerland.}
\vskip .1in
\centerline{\it ${}^b$ Physics Department, McGill University}
\centerline{\it 3600 University St., Montr\'eal, Qu\'ebec,  Canada, H3A 2T8.}
\vskip .1in
\centerline{\it ${}^c$ Department of Physics, University of Maryland}
\centerline{\it College Park, Maryland, USA, 20742.}
\endauthors
%\vskip .3in
\abstract
We investigate the possibility of producing neutrinoless double beta decay
{\it without} having an electron neutrino with a mass in the vicinity
of 1 eV. We do so by having a much lighter electron neutrino mix with a much
heavier ($m \gsim 1 \GeV$) sterile neutrino.  We study the constraints on the
masses and mixings of such heavy sterile neutrinos from existing laboratory,
astrophysical and cosmological information, and discuss the properties it would
require in order to produce a detectable signal in current searches for
neutrinoless double beta decay.
\endabstract

% Main text ----------------------------------------------------------------

\vfill\eject
\section{Introduction}

Our ultimate understanding of the universe relies on two crucial ingredients:
the nature of the constituents (or building blocks), and the nature of the forces
through which they interact. Experience to date indicates that spin-half
fermions are the ultimate constituents, whereas forces arise from the exchange
of gauge bosons associated with local symmetries, and Higgs bosons needed to
break these symmetries. The highly successful Standard Model (SM) is based on
45 chiral fermions (15 for each generation consisting of twelve coloured quark
states and three leptonic ones). Of these,
 the neutrinos have the unique property
of being electrically neutral.

\ref\seesaw{M. Gell-Mann, P. Ramond and R. Slansky, in {\it Supergravity},
ed. by F. van Nieuwenhuizen and D. Freedman (Amsterdam, North Holland,
1979) 315; \bk
T. Yanagida, in the Proceedings of the Workshop on {\it Unified Theory and
Baryon Number in the Universe}, ed. by O. Sawada and A. Sugamoto (KEK,
Tsukuba, 1979) 95;\bk
R.N. Mohapatra and G. Senjanovi\'c, \prl{44}{80}{912}.}
A necessary part of the exploration of physics
beyond the SM is the study of new chiral fermions and their properties. Apart
from sheer curiosity regarding their existence, there are often very good
physical
motivations for postulating such particles. For instance, it is by now well known
that if neutrinos have a mass, the smallness of this mass is easily understood if
there is an additional heavy, isosinglet Majorana neutrino which mixes with the
known ones \via\ the so-called see-saw mechanism \seesaw. Similarly, there also
exist quite good reasons which would motivate the existence of new charged
fermions not present in the standard model. An important area of investigation
in particle physics now is the study of the constraints that may be inferred
on the properties of such new fermions using the existing data, as well as the
identification of new experiments which can further constrain these properties.

\ref\lthree{The L3 Collaboration, \plb{316}{93}{427};
S. Shevchenko, preprint CALT-68-1939 (1993).}
\ref\Gonau{M. Gonau, C. Leung and J. Rosner, \prd{29}{84}{2539}; F.J. Gilman,
{\it Comments\ Nucl.\ Part.\ Phys.\ }{\bf 16 }{(1986) 231}.}
\ref\Dittmar{M. Dittmar, M.C. Gonzalez-Garcia, A. Santamaria and J.W.F.
Valle, \npb{332}{90}{1}.}
\ref\apostolos{A. Pilaftsis, \prd{49}{93}{2398}.}
\ref\bigfit{C.P. Burgess, S. Godfrey, H. K\"onig, D. London and I. Maksymyk,
\prd{49}{94}{6115}.}
\ref\Chang{L.N. Chang, D. Ng and J. Ng, preprint TRI-PP-94-1.}
Since charged fermions cannot be singlets under the SM gauge group, the
LEP data on the $Z$-width already imposes very stringent limits on their
masses, \ie\ $m_\ssf \gsim 45 \GeV$. However, no such direct constraints need
apply for SM-singlet particles, unless they mix strongly with the known
neutrinos. There are, however, many indirect limits on such particles
\lthree, \Gonau, \Dittmar, \apostolos, \bigfit, \Chang\
and our goal in this paper is to explore
these in the context of a specific model and indicate how future
neutrinoless double beta decay ($\bbzn$) experiments can probe the existence of
these particles in interesting ranges of masses and mixing angles.

\ref\moha{R. N. Mohapatra, \prd{34}{86}{3457}; R. N. Mohapatra
and J. D. Vergados, \prl{47}{81}{1713}; J. Schecter and J. W. F. Valle,
\prd{25}{82}{2951}.}
\ref\cancellation{C.N. Leung and S.T. Petcov, \plb{145}{84}{416}.}
\ref\boris{For an excellent discussion of Majorana neutrinos
and their connection to $\beta\beta_{0\nu}$ decay,see B. Kayser,
F. Gibra-Debu and F. Perrier,
{\it Physics of Massive Neutrinos}, ( World Scientific, 1989).}
One motivation for examining the particular type of models we consider here
is that they contradict the maxim\foot\alternatives{ 
Alternative scenarios have also been explored for having observable 
$\ss \beta\beta_{0\nu}$ decay without requiring $\ss m_{\nu_e}$ to
be in the eV range \moha. Our approach here is more similar to, and 
updates, the framework of Ref.~\cancellation.}\ that the observation of
neutrinoless double-beta decay would demonstrate the existence of a mass
for the electron neutrino in the range of 1 eV \boris. In the models we study, all
of the neutrinos are either much lighter than, or much heavier than the
eV mass region. Neutrinoless double beta decay proceeds in these models
through the virtual exchange of the heavier (\ie\ GeV-scale or higher)
neutrinos. We are able to find phenomenologically acceptable
masses and mixings for such a model that are consistent with an observable
signal for double beta decay.

\section{The Model}

We restrict ourselves for simplicity to supplementing the SM with two sterile
neutrinos, $N_\pm$, which mix only with the SM neutrinos of the first generation.
The reasons for adding two sterile neutrinos rather than one are twofold:
(a) the case of one extra sterile neutrino falls into a subclass of models
which use the see-saw mechanism, and which have been discussed
elsewhere \Gonau, \Dittmar;
and (b) in models with one sterile neutrino,
 in the limit $m_{\nu_e} \to 0$, the
sterile neutrino completely decouples and becomes invisible, whereas in models
with two (or more) sterile neutrinos, the situation completely changes, and the
sterile neutrinos can mix appreciably even in the limit $m_{\nu_e} \to 0$,
and so can be potentially visible in many processes \apostolos.

The interaction lagrangian involving $N_\pm$ can be written as follows:
\label\lsum
\eq
\Scl = \Scl_0 + \Scl_1,
\eeq
where $\Scl_0$ conserves lepton number and $\Scl_1$ violates it. (We assign
$L(N_\pm) = \pm 1$ to the left-handed parts of $N_\pm$.) The most general
renormalizable couplings and masses are:
\label\linv
\eq
\Scl_0 = - m \, \Nbar_+ \Pl N_- - \lambdae \, (\Lbar_e \Pl N_-) \, H +  \hc ,
\eeq
and
\label\lninv
\eq
\Scl_1 = - {\mu_+ \over 2} \; \Nbar_+ \Pl N_+ - {\mu_- \over 2} \; \Nbar_- \Pl
N_- - \twlambdae \,  (\Lbar_e \Pl N_+) \, H +  \hc ,
\eeq
where $H$ is the usual SM Higgs doublet, and $L_e = {\nu_e \choose e}$ is the
first-generation lepton doublet. After the SM gauge symmetry breaking --- with
$\Avg{H} = v = 174$ GeV ---  one gets the following most general mass term
relating the left-handed states $\nue$, $N_+$ and $N_-$:
\label\lhmasses
\eq
\Scl_m = - \hf \; \pmatrix{ \ol{\nu}_e \cr \Nbar_+ \cr \Nbar_- \cr}^\sst
\pmatrix{ 0 & \twlambdae v & \lambdae v \cr
\twlambdae v & \mu_+ & m \cr \lambdae v & m & \mu_- \cr}
\pmatrix{ \nue \cr N_+ \cr N_- \cr} .
\eeq
We assume for simplicity, in writing eq.~\lhmasses, that the elements of  the
mass matrix are real. 
This mass matrix has three nonzero eigenvalues, one of which is the physical
electron neutrino, $\nue'$. Since the direct laboratory bounds on the mass of a
dominantly electron neutrino are quite low, we must ask under what circumstances
a massless neutrino can emerge from this matrix. There are two possible regimes:

\item{(1)}
$\twlambdae = \mu_+ = 0$.
The mass eigenstates can in this case be written as:
\label\evecone
\eq
\pmatrix{ \nue' \cr N_+' \cr N_-' \cr} = \pmatrix{ \csone & - \snone & 0 \cr
\snone \cstwo & \csone \cstwo &  -\sntwo \cr i\snone \sntwo &  i\csone  \sntwo
& i\cstwo \cr} \pmatrix{ \nue \cr N_+ \cr N_-  \cr} ,
\eeq
where $\cs i$ ($\sn i$) denotes $\cos \theta_i$ ($\sin \theta_i$) while
\label\tanforms
\eq
%\tan \theta_1 = - \, { \lambdae v \over m} ; \qquad \hbox{and} \qquad
%\tan2\theta_2 = -2 \, { \sqrt{ m^2 + \lambdae^2 v^2 } \over \mu_- }.
\tan \theta_1 =  { \lambdae v \over m} ; \qquad \hbox{and} \qquad
\tan2\theta_2 = 2 \, { \sqrt{ m^2 + \lambdae^2 v^2 } \over \mu_- }.
\eeq
Finally, the factors of `$i$' in the last row of eq.~\evecone\ come from the
chiral rotation that is required to ensure that all of the entries in the final
mass matrix are positive. The corresponding masses are:
\label\evalone
\eq
M_{\nue'} = 0 , \qquad \hbox{and} \qquad M_{\npmpm} = \hf \left[
\sqrt{ \mu_-^2 + 4 ( m^2 + \lambdae^2 v^2 ) }  \mp \mu_- \right],
\eeq
Notice that in the limit $\mu_- \to 0$, lepton number is exactly conserved,
and $N'_\pm$ forms a degenerate Dirac pair. In this case all lepton-number violating
processes, such as $\bbzn$ decay, are forbidden.

\item{(2)}
$\lambdae = \mu_- = 0$.
This alternative is identical to case (1) above, but with the two sterile states
$N_\pm$ interchanged. The physical implications of this case are therefore
identical to case (1), and we need not further pursue this alternative
separately. We henceforth exclusively focus on mass-matrix
parameters that are in the vicinity of case (1).

For later purposes it is useful to define a dimensionless parameter, $\epsilon$,
which measures the strength of the lepton-number violation in the mass matrix:
\label\epsdef
\eq
\epsilon = {\mu_- \over  \sqrt{ m^2 + \lambdae^2 v^2}},
\eeq
The ratio of the nonzero mass eigenvalues have a simple expression in terms of
this parameter:
\label\epsratio
\eq
{M_{\npmm} \over M_{\npmp} } = { \sqrt{ 1 + 4/\epsilon^2} + 1 \over
 \sqrt{ 1 + 4/\epsilon^2} - 1} = \cot^2\theta_2 \ .
\eeq
In the lepton-conserving limit, $\epsilon \ll 1$, $M_{\npmm} \simeq M_{\npmp}$
as would be expected for pseudo-Dirac particles. In this case the mass eigenstates, 
$N'_\pm$, have opposite CP properties. For $\epsilon \gg 1$ we instead have 
$M_{\nm} / M_{\np} \sim \epsilon^2$, so that in this limit $N'_-$ is much heavier 
than $N'_+$. 

We could now turn on small non-vanishing values for $\twlambdae$ and $\mu_+$,
so that the electron neutrino acquires a small nonzero mass. However, we are
interested in the situation for which $M_{\nue} \ll 1 \eV$ and yet for which
there are nevertheless potentially observable contributions to $\bbzn$ decay.
We therefore keep $\twlambdae$ and $\mu_+$ negligibly small in what follows,
although we return to the naturalness of this choice in the following section.
We envision the possibility that the $N_\pm'$ masses span a wide range of
possible
values, starting from the MeV up to the GeV range. Clearly, the properties of
such
particles are highly constrained by known laboratory, astrophysical and
cosmological
information. We discuss the ranges of masses and mixings that are not already
ruled out by these constraints and see if $\bbzn$ decay can be observable in the
allowed range.

\section{Contributions to $\bbzn$ Decay}

As may be seen from eq.~\evecone, the heavy sterile neutrinos acquire
charged-current weak interactions through their mixing with the electroweak
eigenstate $\nue$, resulting in Kobayashi-Maskawa-type mixing angles
\label\mixangles
\eq
U_{e\nue'} = \csone, \qquad U_{e\npmp} = \snone \cstwo ,
\qquad U_{e\npmm} = -i \snone \sntwo .
\eeq
\ref\jimandi{C.P. Burgess and J.M. Cline, \plb{289}{93}{141}; \prd{49}{94}{5925}.}
Provided that $\mu_- \ne 0$, so that lepton number is broken, neutrinoless
double beta decay arises in this model via the exchange of the three neutrino
states. The differential decay rate for this decay between two $0^+$ nuclei
can be written in the following simple form \jimandi:
\label\bbznrate
\eq
{d \Gamma \over d\varepsilon_1 d\varepsilon_2} = { \GF^4 \cos^4 \theta_c
\over 2 \pi^3} \;  | \Scw |^2 \; \delta(Q - \varepsilon_1 - \varepsilon_2) \;
[ \varepsilon_1 p_1 F(\varepsilon_1) ] [\varepsilon_2 p_2 F(\varepsilon_2) ].
\eeq
where $\GF$ is Fermi's constant; $\theta_c$ is the Cabbibo angle which governs
the strength of the hadronic charged-current; $\varepsilon_i$ and $p_i$ are the
energy
and momentum of each of the final two electrons; $Q = M(Z,A) - M(Z+2,A) -
2m_e$ is their endpoint energy --- typically several MeV; and $F(\varepsilon)$
is the Fermi function which describes the distortion of the electron spectrum
due to the nuclear charge. $Z$ and $A$ represent the charge and mass
number of the initial nucleus.  For our present purposes it is convenient to
work with analytic expressions for the total decay rate, which we can obtain
if we make some simplifying assumptions, which are sufficiently
accurate for the estimates in this paper. In performing the phase-space integrals
we therefore neglect: ($i$) the electron mass, and ($ii$) 
the Coulomb-distortion factor, $F(\varepsilon)$.

The dependence on the neutrino masses, in eq.~\bbznrate, lies in the 
quantity $\Scw$, which is given explicitly by:
\label\wdef
\eq
\Scw = \sum_i U_{ei}^2 \; m_i \int {d^4p \over (2\pi)^4} \;
   \left( { w  \over p^2 - m_i^2 + i\varepsilon} \right).
\eeq
The sum here is over all three neutrino species.  $w = w(p_0, |\bfp|)$
represents a particular Lorentz-invariant combination of form factors
describing the nuclear matrix element of the two hadronic charged currents.
\ref\Halprin{A. Halprin, P. Minkowski, H. Primakoff and P. Rosen,
\prd{13}{76}{2567}.}
\ref\Haxton{W. Haxton and G. Stephenson, {\it Prog. in Particle and Nucl. 
Physics} {\bf 12}, 409 (1984);
 J. Vergados, {\it Phys. Rep.} {\bf 133}, 1 (1986);
M. Doi, T. Kotani, H. Nishiura and E. Takasugi, {\it Prog. Theor. Phys.}, 
{\it Prog. Theor. Phys. Suppl.} {\bf 83}, 1 (1985).}
\ref\KK{H. Klapdor-Kleingrothaus \etal, Heidelberg Preprint, (1993).}
All of the theoretical uncertainty in the decay rate enters with the
estimating of $w$ within a model of the nucleus. The connection between $\Scw$
as defined here and the usual estimates \Halprin, \Haxton, \KK, based on an
independent-nucleon model of the nucleus within the closure approximation,
is given by \jimandi:
\label\znnmes
\eq
\Scw =  \sum_i {U^2_{ei} m_i \over 4\pi} \; \bra{N'(Z+2,A) }  \sum_{mn}
\tau_m^+\tau_n^+ \; h(\bfr_{nm} ;m_i)
        \left(g_\ssv^2 -g_\ssa^2 \vec\sigma_n \cdot\vec\sigma_m\right)
\ket{N(Z,A)},
\eeq
where $m$ and $n$ run over the labels of the nucleons in the nucleus;
$\tau^+_m$ and $\vec\sigma_m$ are isospin and spin matrices for
the $m$th nucleon; and $g_\ssv$ and $g_\ssa$ are the vector and axial
charged-current couplings of the nucleon. The function $h(\bfr_{mn} ;m_i)$
of the internucleon separation, $\bfr_{mn}$, is the neutrino potential,
which is defined by the following integral:
\label\nupotential
\eq
h(\bfr ; m) = {1\over 2\pi^2} \int d^3\bfp\; {\exp(-i\bfp\cdot\bfr)
\over \omega(\omega+\mu)}; \qquad \omega = \sqrt{ \bfp^2+m^2} ,
\eeq
where $\mu = \hf \; [M(Z,A) +M(Z+2,A) ]$ is the mean excitation energy of
the nucleus.

\ref\bbexp{A. Piepke, TAUP '93, ( to appear in the proceedings);
Older references are: A. Piepke, presented at ICHEP XXVI, Dallas, Aug. 1992;
D. Caldwell \etal, {\it Nucl. Phys.} (Proc. Suppl.) {\bf B13} 1990 547;
\ijmp{4}{89}{1851}; 
M. Moe and P. Vogel, {\it Ann. Rev. of Nucl. and Part. Sc.} (to appear).}
\ref\bbtheory{T. Tomoda, {\it Rep. Prog. Phys.} {\bf 54} (1991) 53.}
For the present purposes, however, we need not use the detailed matrix-element
machinery, as quite good analytic results can be obtained by making the following
simplifying approximation for the functional form for $w$. 
We parameterize $w$ by representing it as a step functions in
energy and momentum: $w(p_0,|\bfp|) \simeq w_0 \, \Theta(p_0 - E_\ssf) \,
\Theta(|\bfp| -  \pf)$. Here $\pf$ represents the nucleon Fermi momentum
and $E_\ssf = \pf^2/2m_{\sss N}$ is the corresponding Fermi energy. By
requiring this approximation to reproduce the observed $\beta\beta_{2\nu}$ decay
rates, we find $w_0 \simeq 4 \, \MeV^{-1}$, and by requiring that the present
upper limit on the half life for neutrinoless decay \bbexp\ imply an upper
limit for light neutrino masses \bbtheory\ of $| U_{e\nu}^2 m_\nu | \lsim
2$ eV,\foot\whytwo{We take here a bound which is twice as large as the
usually-quoted limit, since we allow for an uncertainty in the nuclear matrix
elements of a factor of 2.} we find $\pf \simeq 60 \, \MeV$ (and so $E_\ssf \simeq 2$ MeV).

With these choices we obtain the following expression for the total $\bbzn$
decay rate:
\label\psint
\eq
\Gamma(\bbzn) \simeq { \GF^4 Q^5 \cos^4 \theta_c \over 60 \pi^3}  \; | \Scw |^2 ,
\eeq
with $\Scw$ given by an elementary integral:
\label\simpw
\eq \eqalign{
\Scw &\simeq {i w_0 \over 4\pi^3} \sum_i U^2_{ei} m_i \int_0^{E_\ssf} du \left[
\pf - \sqrt{u^2 + m_i^2} \; \tan^{-1}\left({\pf \over \sqrt{u^2 + m_i^2} }
\right)
\right] , \cr
& \approx {i w_0  \over 4\pi^3} \; \sum_i U^2_{ei} m_i  \left\{ E_\ssf \pf
- { \pi E_\ssf^2 \over 4} - {\pi m_i^2 \over 4} \left[ \log \left(
{2 E_\ssf \over m_i} \right) + \hf \right] \right\} + \cdots
\qquad (m_i \ll E_\ssf \ll \pf) , \cr
& \approx {i w_0 \over 4\pi^3} \; \sum_i {U^2_{ei} \, E_\ssf 
\pf^3 \over 3 m_i } + \cdots \qquad\qquad  (E_\ssf \ll \pf  \ll m_i) . \cr}
\eeq

Motivated by the expression for the decay rate due to light neutrinos,
it has become conventional to quote the experimental limit on $\bbzn$ decay
as an upper limit on the `effective' neutrino mass, defined by
$m^{\rm eff}_\nu = \sum_i U^2_{ei} m_i$. As mentioned previously, the
current experimental limit \bbexp\ implies an upper bound
$m^{\rm eff}_\nu \lsim 2 \eV$ \bbtheory. From the above formulae for
$\Gamma(\bbzn)$ we can obtain similar constraints on $U_{ei}$ and
$m_i$ for arbitrary neutrino masses. For the model under study here,
we assume the $\nue'$ mass to be too small to contribute, and so there are
three limiting cases to consider:

\item{(1)}
If the two sterile neutrinos are both light compared to a few MeV, then
one would expect $m^{\rm eff}_\nu =  \snone^2 \left( M_{\npmp}
\cstwo^2 - M_{\npmm} \sntwo^2 \right)$ as the approximate expression for
$m^{\rm eff}_\nu$. However, in the model considered here this quantity
vanishes, as may be seen from eq.~\epsratio. As a result we must
work to sub-leading order in the sterile neutrino masses when approximating
the integral in eq.~\simpw, leading to
\label\llbound
\eq
m^{\rm eff}_\nu=
{\pi \snone^2\over 4E_\ssf\pf} \left| M_{\npmp}^3 \cstwo^2\left[ {1\over 2} +
\log\left( {2E_\ssf \over M_{\npmp}} \right)\right] -
 M_{\npmm}^3 \sntwo^2 \left[ {1\over 2}+\log
\left( {2E_\ssf\over M_{\npmm}} \right)\right] \right|
 \lsim 2 \, \eV.
\eeq

\item{(2)}
If both sterile neutrinos have large masses compared to $\pf  \simeq 60$ MeV, then we
instead find
\label\hhbound
\eq
 {\snone^2 \pf^2 \over 3} \left| {\cstwo^2 \over M_{\npmp}}
- {\sntwo^2 \over M_{\npmm}} \right| \lsim 2 \, \eV.
\eeq

\item{(3)}
Finally, if $M_{\npmp} \ll \pf$ and $M_{\npmm} \gg \pf$, then
\label\lhbound
\eq
 \snone^2 \left| M_{\npmp} \cstwo^2 -
{\sntwo^2 \pf^2 \over 3 M_{\npmm}} \right| \lsim 2 \, \eV.
\eeq

\figure\bbznbounds{This figure plots the region in the $U_{ei} - m_i$
plane (Fig.~1a, (1b) for $N'_+$, ($N'_-$ respectively)) that is allowed by
current $\bbzn$ experiments for various different values of the parameter 
$\epsilon $ as defined in eq.~\epsdef . The area
below the corresponding lines is the allowed region.
The limiting case of very large $\epsilon$ appears as the solid line labelled 
``$\epsilon = \infty $'', which also corresponds to the contribution of a single 
sterile neutrino that mixes with $\nu_e$.}

With these estimates we can determine the masses and couplings of the sterile
neutrinos which are consistent with the present non-observation of $\bbzn$.
The constraints we obtain for $U_{ei}$ and $m_i$ in this way are plotted in
Figs.~\bbznbounds. The area below the curves in this figure represents
the allowed range. To see if a sterile neutrino will make observable
contributions to $\bbzn$ decay, we must see if the range of values which lie
close to those curves are consistent with other constraints.

\subsection{Radiative Corrections}

Our estimate of the $\bbzn$ decay rate in this section assumes a negligible
contribution from the exchange of the very light neutrino mass eigenstate,
$\nue'$. This assumption requires some justification in parts of the
parameter space which we consider here.
We therefore briefly pause to provide this justification.

The neglect of the light neutrino contribution to $\bbzn$ relies on our
decision to choose this mass to be much lighter than 1 eV.  At tree level
this can always be accomplished by choosing the parameters $\twlambdae$
and $\mu_+$ to be sufficiently small. A naturalness problem can arise,
however, if $m_{\nue'}$ is chosen to be too small in comparison with the
parameter $\mu_-$,
which provides the lepton-number violating contribution to the heavy-neutrino
masses.  This is because a nonzero $\mu_-$ generates, through loops,
nonzero contributions to $\twlambdae$ and $\mu_+$, and if these loop-induced
contributions are large enough, then the $\nue'$ exchange graph
can only be neglected in $\bbzn$ if the loop-induced mass is cancelled
by the tree-level term. A naturalness problem
arises if  the required cancellation becomes implausibly precise.

More quantitatively, we imagine our model to be an effective theory which
is obtained after some unknown physics above some scale $\Lambda$ has
been  integrated out. We then use the renormalization group to run the
couplings in this effective theory down from the scale $\Lambda$ to
the much lower energies that are relevant for $\bb$ decay. In this way
we can compute the contribution to the $\nue'$ mass which
is produced by the mixing between the parameters $\twlambdae$ and
$\mu_+$, and $\mu_-$ as they are run down from
the scale $\Lambda$. We regard the theory to be natural if these
contributions to the $\nue'$ mass are not larger than, say,  $O(1 \, \eV)$,
and so do not need to be carefully cancelled by a `bare' contribution
to $m_{\nue'}$ at the scale $\Lambda$.

\figure\loopgraph{The Feynman graph which dominates the
renormalization-group mixing of $\mu_-$ with $\twlambdae$ and
$\mu_+$.}

The dominant graph to consider is that of Fig.~\loopgraph, in which a SM Higgs
scalar is emitted and absorbed by the light neutrino state. Its contribution
to the light-neutrino mass is, in order of magnitude:
\label\loopmass
\eq
\delta M_{\nue'}(\mu) \sim \left( {\lambdae \over 4 \pi} \right)^2 \mu_- \, \log \left(
{ \Lambda^2 \over \mu^2} \right).
\eeq
\figure\phenconstr{This figure summarizes the phenomenological constraints as
discussed in section 4. The region above the solid line is ruled out by the 
various laboratory bounds. The region below and to the 
left of the dashed line labelled $SN$
is excluded by the observations of SN1987a. The dashed line 
labelled $NN$ depicts the naturalness bound as discussed in section 3 
(Eq.~\loopmass ) for $\epsilon = 0.1$. In the region
to the right and above this line fine tuning is required.
Finally the dash-dotted curves labelled $a$, $b$ and  $c$ represent the 
nucleosynthesis bounds. The lifetime of a  sterile neutrino is
less than $0.1$ sec to the right of curve $a$. 
A particle decouples after
having become non relativistic in the region to the right of line $b$, but it 
will nevertheless decouple at or above a temperature of $\sim 100$ MeV to the 
right and below line $c$. Line $a$ thus represents the only relevant nucleosynthesis 
bound in our case, and the region to the right of line $a$ and below the solid line is
allowed.}
\figure\parameterspace{We plot here the region of the parameter space
in the $\epsilon - M_\npmp$ plane which is obtained by requiring masses and 
mixing angles which yield a $\bbzn$ signal close to observability, as depicted 
by the various lines in Fig.~\bbznbounds, together with the various 
phenomenological bounds displayed in Fig.~\phenconstr. The allowed area 
is marked by shading, and extends upwards beyond the region depicted in the figure
towards higher values of $\epsilon $ without changing the mass range.
The darker area represents the part of the parameter space in which 
the smallness of the $\nue'$ mass is explained in a way which is technically natural 
in the sense explained in section (3.1) of the text. In the lightly-shaded region 
a finetuning of $\twlambdae$ and $\mu_+$ is required in order to keep the
$\nue'$ mass below 1 eV. Notice that, since $M_\npmm \ge M_\npmp$,
the analogous figure in the $\epsilon - M_\npmm$ plane would be shifted to 
the right according to eq.~\epsratio .}
For numerical purposes we take the logarithm in this expression to be
unity. Requiring the rest of the result to be smaller than $O(1 \, \eV)$ then
produces the bound labelled $NN$ in Fig. \phenconstr .

\section{Phenomenological Constraints}

We now turn to the exploration of the other constraints on this model. We
consider
in turn the limits coming from ($i$) supernova SN1987a, ($ii$)
nucleosynthesis, and ($iii$) laboratory limits.

\vfill\eject
\subsection{Supernova SN1987a}

\ref\Barbieri{R. Barbieri and R.N. Mohapatra, \prd{39}{89}{1229};
G. Raffelt and D. Seckel, \prl{60}{89}{1793}.}
The first constraint we consider is for low-mass sterile leptons
(`low-mass' here means masses smaller than $\sim 50$ MeV) that mix 
significantly with $\nue$. Any such particle can be produced in the core 
of a supernova, where temperatures are typically of order $T_{\sss SN} 
\simeq (60 - 70)$ MeV. This must be forbidden since otherwise the supernova
would cool too much to be in agreement with the observations of SN1987a.
The analysis we require is very similar to the case of  right-handed neutrinos 
discussed in Ref.~\Barbieri, where it was shown that these considerations
lead to the following bounds
\label\snbound
\eq
 3 \times 10^{-2} \lsim |U_{ei}|,\; {\rm or\ else}\;\; |U_{ei}|
 \lsim 10^{-5}.
\eeq
The lower bound comes because for sufficiently strong mixing, the
produced sterile neutrinos get trapped in the supernova and so they 
do not provide a mechanism for cooling too quickly. The upper bound
comes from the requirement that not too many sterile neutrinos
be radiated by the ordinary particles in the supernova. 
These bounds are displayed in
Fig.~\phenconstr\ by the vertical and horizontal lines labelled $SN$. 
\ref\NEMO{See the NEMO III Proposal, (1994).}
The vertical line is due to this bound being
independent of the neutrino mass, provided only that this mass
is much smaller than $T_{\sss SN}$. For the models of interest here, the 
upper bounds we obtain in this way for $M_i$ and $|U_{ei}|$ imply
that the effective mass, $m^{\rm eff}_\nu$, which appears in $\bbzn$ decay
can be at most $m^{\rm eff}_\nu \lsim 5 \times 10^{-3}$ eV. This is well beyond
the reach of present- and next-generation \NEMO\ $\bb$-decay experiments.
It is, however, worth noting that for $M_{\npmpm} \gsim 50$ MeV, this bound is
ineffective, and so SN1987a cannot rule out a significant contribution of 
sterile neutrinos to the $\bbzn$ decay rate, provided that these neutrinos are
in this larger-mass regime.

We therefore now turn to the constraints on $N'_\pm$ which apply if the masses
are 50 MeV or higher.

\subsection{Cosmological Constraints}

\ref\BBN{T. Walker \etal, {\it Ap. J.} {\bf 376}, 51(1991). }
One of the major triumphs of the standard hot big-bang model of 
cosmology is its
ability to naturally explain the primordial abundance of the light
elements --- ${}^4$He, ${}^7$Li, ${}^2$H and ${}^3$He --- using the standard
model (SM) of electroweak interactions \BBN. This, in turn, implies
stringent bounds on any new physics beyond the SM that involves
weakly-interacting particles. In particular, in the model of present interest,
the sterile neutrinos can upset the success of the nucleosynthesis discussion
unless their masses and mixings are suitably constrained.

The basic condition is to ensure 
that the energy density at the nucleosynthesis temperature, 
$T_{\sss BBN} \simeq 1$ MeV, due to the sterile neutrinos
is much less than that of an ordinary neutrino species. For 
sterile neutrinos that are much lighter than 1 MeV, this can be
arranged simply by having them decouple early enough for
their energy density to be diluted by the reheating of 
ordinary matter, such as in the QCD phase transition. 
But keeping in mind the supernova constraint
of the previous section, the sterile neutrinos in the model of interest
here satisfy $M \gg T_{\sss BBN}$,
and so even if they decouple sufficiently early, their relic energy density at 
nucleosynthesis will nevertheless dominate that of an ordinary neutrino species 
unless their lifetime is shorter than  0.1 sec. Assuming decays through
the charged-current weak interactions, this implies:
\label\bdlifeone
\eq
\left( {M_{\npmpm} \over \GeV} \right)^5 |U_{ei}|^2 \gsim 3.6 \times 10^{-11}  .
\eeq

Suppose, first, that the sterile neutrino is relativistic at the time that it 
decouples: $T_\ssd \gsim M_{\npmpm}$. This is the case if
\label\bdone
\eq
\left( {M_{\npmpm} \over \GeV} \right)^3 |U_{ei}|^2 \lsim 3 \times 10^{-8} .
\eeq
Both of these conditions must be satisfied by both of the heavy particles, and are
plotted as curves $a$ and $b$ in Fig.~\phenconstr. BBN demands that any particle 
which lies to the left of curve $b$ (\ie\ decouples relativistically) must also 
lie to the right of curve $a$. 
Notice that these conditions together drive one, in the model of current interest,
into a regime which is excluded by the laboratory bounds considered in the
following section, provided one asks for a $\bbzn$ rate close to observability
at the same time, and so the relativistic-decoupling scenario is not relevant
in our case. 

Next suppose that the sterile neutrinos decouple nonrelativistically,
$T_\ssd \lsim M_{\npmpm}$, and so
lie to the right of curve $b$ in Fig.~\phenconstr. 
In this case their number density is 
exponentially suppressed by the 
Boltzmann factor. This dilution then makes them decouple
earlier than one would otherwise expect without the Boltzmann suppression.
The decoupling temperature, $T_\ssd$, is therefore given by
\label\bdtwo
\eq
{ M_{\npm}^2 \GF^2 |U_{ei}|^2 \over \pi} \; ( M_{\npm} T_\ssd)^{3/2} 
e^{-M_{\npm}/T_\ssd} \lsim g_\star^{1/2} \; {T_\ssd^2 \over M_{pl} }.
\eeq

In the case of present interest the sterile neutrinos will always
decouple at or above a temperature of $\sim 100 MeV$, i.e. lie to the 
right or below curve $c$ in Fig.~\phenconstr, provided $\bbzn$ occurs
at or close to an observable level. This in turn means that their
energy density at nucleosynthesis will only be sufficiently small
if they, again, decay fast enough, eq.~\bdlifeone.

To summarize, having observable $\bbzn$ in this model implies that the 
heavy sterile neutrinos decouple only after having become non relativistic,
but still early enough that conflict with standard Big-Bang Nucleosynthesis can be
avoided, provided they decay fast enough.

\subsection{Laboratory Limits}

A wide range of experiments constrain the properties of isosinglet heavy leptons
\lthree, \Gonau, \Dittmar, \apostolos, \bigfit, \Chang. They can be
classified into two main categories,
according to whether the bound is obtained from precision measurements
on the $Z$ resonance, or from experiments at lower energies.

\ref\brit{D.I. Britton \etal, \prd{46}{92}{R885}; \prl{68}{92}{3000}.}
\ref\kaonbound{ R. Shrock, \plb{96}{80}{159};
T. Yamazaki \etal, in Proc. XIth Intern. Conf. on Neutrino
Physics and Astrophysics, K. Kleinknecht and E.A. Paschos (World Scientific,
Singapore, 1984) p. 183.}

There are two main types of low-energy experiments which limit the
properties of sterile neutrinos that mix with the electron neutrino, such as
for the model considered here. One type obtains its bound from the decay
rate and the electron spectrum of  the two-body decay of kaons and pions at rest.
For example, for neutrinos in the mass range between 1 and 100 MeV,
the measured $\Gamma(\pi\to e\nu)/\Gamma(\pi\to \mu\nu)$ rate
provides a mass-dependent bound on $|U_{ei}|$. For a 1~MeV neutrino
the bound is $|U_{ei}|^2 < 10^{-3}$ at the 90\% C.L., whereas the respective
bounds for a 10~MeV and a 50~MeV neutrino are $10^{-5}$ and $5 \times
10^{-7}$ \brit. Similar searches for a nonstandard component to $K \to e \nu$
\kaonbound\ extend this limit up to sterile-neutrino masses of 350 MeV.

\ref\beamdump{G. Bernardi {\it et.al.}, \plb{203}{88}{332};
J. Dorenbosch {\it et.al.}, \plb{166}{86}{473};
R.C. Ball {\it et.al.}, {preprint UM HE 85-09}, (1985) unpublished;
A.M. Cooper-Sarkar {\it et.al.}, \plb{160}{85}{207}.}

Even stronger limits can be obtained from beam-dump experiments
provided that the heavy neutral leptons can decay appreciably through
their charged-current interactions. These can constrain sterile-neutrino
masses up to $\sim 2\ GeV$, with a sensitivity of  $|U_{ei}|^2 \lsim 10^{-7}$.
For masses below $\sim 0.5 \, \GeV$ the bound becomes as good as
$|U_{ei}|^2 \sim 0.5 \times 10 ^{-9}$ \beamdump.
(For a more detailed discussion see \Gonau).

For neutrinos with masses that are more than a few GeV, measurements at
the $Z$ pole provide the strongest limits. For $M_\npmpm < M_\ssz$, the best
bounds come from the nonobservation of the decay of a $Z$ into a sterile
and a standard neutrino, $Z \to N \bar{\nu } \to W^* e \bar{\nu}$, 
with the subsequent decay of the sterile neutrino through a virtual boson,
$W^*$. The bound obtained in
this way is $|U_{ei}|^2< 7\times 10^{-5}$  \lthree, \Dittmar, \Chang.

The above bounds do not apply if the heavy singlet neutrino is
heavier than $M_Z$. In this case there are two types of  bounds to
consider, which arise due to the reduction of the couplings of the ordinary
neutrinos to the $W$ and $Z$ bosons due to their admixture with the new
sterile neutrinos. The reduction in the effective
couplings to the $Z$, result in a reduction of the $Z$'s invisible width.
This leads to the bound: $|U_{ei}|^2<2.7\times 10^{-2}$  for isosinglet 
masses above $90\ GeV$ \bigfit, \Chang. A stronger bound arises however
from the reduction of the $W$ couplings, which potentially show up 
as a failure of lepton universality in low-energy weak decays, as well
as affecting precision electroweak measurements through their influence
on the experimental value of Fermi's constant, $G_\ssf$, that is inferred
from muon decay. This implies the bound : $|U_{ei}|^2<5.6\times 10^{-3}$
$(2\sigma )$ for isosinglet masses above $90\ GeV$ \bigfit .

All the phenomenological constraints discussed in this section have been
summarized
in Fig.~\phenconstr . When contrasted with the masses and mixing angles required
for a $\bbzn $ signal close to observability in the model discussed here
(Figs.~\bbznbounds ) they yield the
region of parameter space depicted in Fig.~\parameterspace .

\section{Conclusions}

Our purpose here has been to determine which kinds of heavy sterile neutrinos
can contribute appreciably to $\bbzn$ decays, and to explore the constraints
which such particles must satisfy due to present laboratory and astrophysical
information. Part of our motivation for doing so has been to provide
an example of a theory in which this decay can proceed without requiring
the existence of a light neutrino with a mass in the vicinity of 1 eV.
As an existence proof for theories of this type, we display here a model
which does so. It does so by producing observable $\bbzn$
purely through the exchange of a sterile neutrino having a mass in the
GeV range. This runs contrary to the usual
expectation that the observation
of $\bbzn $ must indicate the existence of a majorana mass for the electron
neutrino in the $eV$ range.

The model we consider is reasonably simple, supplementing the standard
model only by two new left-handed neutrino states.
We find that requiring the model to be consistent with all astrophysical
and laboratory limits, as well as with an observable
$\bbzn$ signal, constrains the couplings and masses of the new neutrinos
to lie in a limited region of parameter space. The mass range that is
favoured by these bounds, as well as naturalness considerations,
is 1 -- 10 GeV. For these masses, the couplings that are required
to produce an observable $\bbzn$ signal are roughly
$|U_{ei}|^2 \sim 10^{-5}$. Such parameters place such
a sterile neutrino close to the current limits of detection at LEP,
where they can be searched for through the decay $Z \to
N \nue $, with the subsequent charged-current decay of the
sterile neutrino, $N$, into quarks and leptons. This shows
how LEP results can be used to help diagnose the implications
of a potential $\bbzn$ signal, and illustrates the rich interplay that
is possible between low- and high-energy experiments.

\bigskip

\centerline{\bf Acknowledgments}

\bigskip

We would like to acknowledge helpful conversations with K. Babu,
as well as research support from N.S.E.R.C.\ of
Canada, les Fonds F.C.A.R.\ du Qu\'ebec, the U.S. National Science Foundation
(grant PHY-9119745) and the Swiss National Foundation.

\listrefs

\figurecaptions

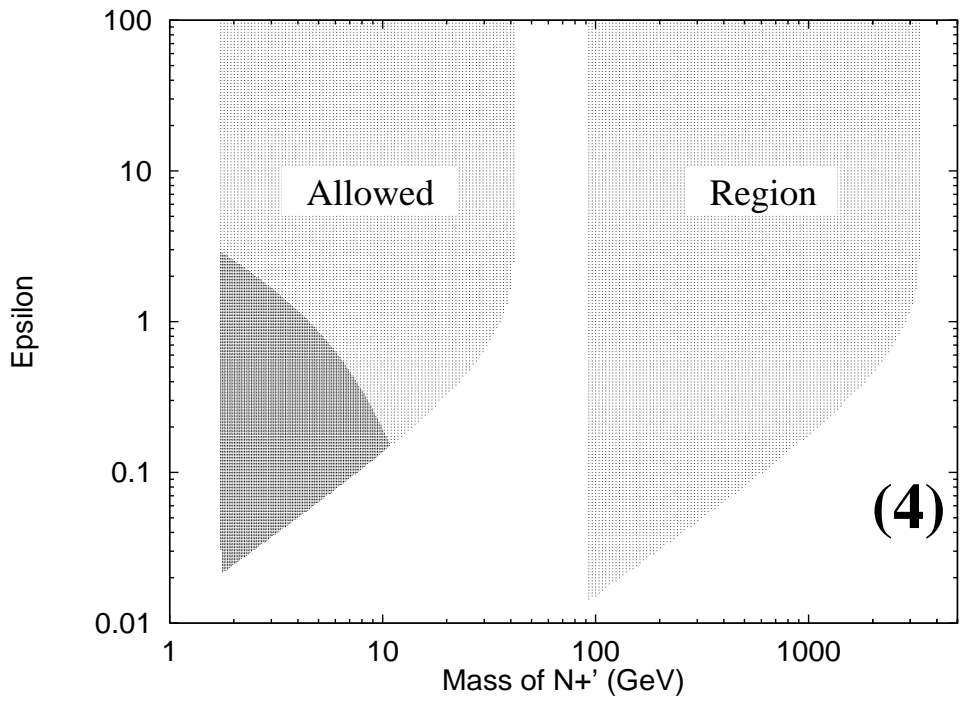
\bye